\documentclass[12pt]{article}

\newcommand{\be}{\begin{equation}}
\newcommand{\ee}{\end{equation}}
\newcommand{\bea}{\begin{eqnarray}}
\newcommand{\eea}{\end{eqnarray}}
\newcommand{\nn}{\nonumber\\}

\newcommand{\rhs}{{\it r.h.s.\/}} 
\newcommand{\ie}{{\it i.e.\/}}

\renewcommand{\d}{\partial}
\newcommand{\la}{\langle}
\newcommand{\ra}{\rangle}

\newcommand{\sla}{\not \!}

\makeatletter 
\def\secteqno{\@addtoreset{equation}{section}%
\def\theequation{\thesection.\arabic{equation}}}

\begin{document}


%
\renewcommand{\thefootnote}{\fnsymbol{footnote}}

{\hfill \parbox{4cm}{ 
        MIT-CTP-3081 \\ 
        hep-th/0102021 }} 

\bigskip\bigskip

\begin{center} \large \bf 
Physical (ir)relevance of ambiguities to \\ Lorentz and CPT violation in
QED 
\end{center}

\bigskip\bigskip

\centerline{
Manuel P\'{e}rez-Victoria\footnote[1]{manolo@lns.mit.edu} }

\bigskip
\centerline{ \it Center for Theoretical Physics}
\centerline{ \it Massachusetts Institute of Technology}
\centerline{ \it Cambridge, {\rm MA}  02139}
\bigskip
\bigskip

\renewcommand{\thefootnote}{\arabic{footnote}}

\centerline{\bf Abstract} 
\begin{center}
\parbox{11cm}{We discuss the regularization and renormalization of QED
with Lorentz and CPT violation, and argue that the coefficient of the
Chern-Simons term is an independent parameter not determined by gauge
invariance. We also study these issues in a model with spontaneous
breaking of Lorentz and CPT symmetries and find an
explicit relation with the ABJ anomaly. This explains the observed
convergence of the induced Chern-Simons term.}
\end{center}

\medskip


\noindent An effective description of the Standard Model incorporating
possible 
Lorentz and CPT non-invariant terms has been developed by Colladay and
Kosteleck\'y~\cite{Colladay:1998fq,Colladay:1997iz,Kostelecky:1999rh,
Colladay:2000jh,Colladay:1998pc}   
and by Coleman and Glashow \cite{Coleman:1997xq,Coleman:1999ti}. One 
particular Lorentz and CPT breaking term involves the photon and has the
Chern-Simons form
\be
{\cal L}^{\mbox{\tiny CS}} = \frac{1}{2}
\epsilon^{\mu\nu\rho\sigma}  \kappa_\mu A_\nu F_{\rho\sigma}
\label{CSterm} \, , 
\ee
where $\kappa_\mu$ is a constant vector that produces a privileged
direction in space-time. ${\cal L}^{\mbox{\tiny CS}}$ is not gauge
invariant but the corresponding term in the action is. This
Chern-Simons term has been well studied in the
literature~\cite{Jackiw:1981kv,Andrianov:1995qv,Andrianov:1998wj,
Andrianov:1999ay, Jackiw:1999js,Adam:2001ma}  
and it has been shown that a timelike 
$\kappa_\mu$ produces 
instabilities and has problems with causality or unitarity. A
consistent quantization of the theory seems possible, nevertheless, for
a spacelike $\kappa_\mu$~\cite{Adam:2001ma}. On the other hand, astrophysical
observations put very stringent
experimental limits on this parameter, indicating a vanishing
$\kappa_\mu$~\cite{Carroll:1990vb}. There 
has been some recent interest and
controversy [15--30]
about
the possibility that this Chern-Simons term be induced by radiative
corrections arising from a CPT and Lorentz violating term in the
fermionic sector: 
\be
{\cal L}^b =  b_\mu \bar{\psi} \gamma_5 \gamma^\mu
\psi \, , 
\ee
with $b_\mu$ a constant vector. If this were the case, it seems that the
strong bounds on   
$\kappa_\mu$ would translate into strong bounds for those non-invariant
terms. The mentioned controversy arises from the fact
that the calculations give an ambiguous finite radiative
correction. Throughout this paper, ``ambiguous'' means regularization 
dependent. Some papers claim that a particular method gives the
correct result~\cite{Chan:1999nk} and others argue that gauge
invariance requires a vanishing induced term
\cite{Coleman:1999ti,Bonneau:2001ai,Battistel:2000ms}. The purpose of
this letter is to try to clarify the situation and to discuss whether
(and to what extent) a Lorentz and CPT violating fermion sector is
excluded by the astrophysical limits on $\kappa_\mu$. 
First we argue that the radiative correction to the Chern-Simons term
is not fixed by gauge invariance but by normalization conditions. Then
we study a simple generalization of this theory in which the
induced term is directly connected to the ABJ
anomaly~\cite{Bell:1969ts,Adler:1969gk}.

The correction to ${\cal L}^{\mbox{\tiny CS}}$ in the effective
action can be extracted from the zero momentum limit of the parity-odd
part of the vacuum polarization:
\be
(\Delta \kappa)_\mu = -\frac{1}{2} b_\mu K(0)\, ,
\ee
where
\be
\Pi_{\mbox{\tiny odd}}^{\mu\nu}(p)=\epsilon^{\mu\nu\rho\sigma}
b_\rho p_\sigma K(p)\, .  
\ee
$\Pi_{\mbox{\tiny odd}}^{\mu\nu}(p)$ is superficially linearly
divergent by power 
counting. We assume that some well-defined regularization method
is used. One degree of divergence is carried away by the
external momentum, so $K(p)$ is at most logarithmically
divergent. At one loop it turns out that the divergent part in the
integrand is proportional to $(k_\mu k_\nu-1/4 \eta_{\mu\nu} k^2)/k^6$,
where $k$ is the loop momentum. Hence, the integral is finite but
regularization dependent. The structure $x_\mu x_\nu/x^2$ that appears
in non-perturbative
methods~\cite{Perez-Victoria:1999uh,Chung:1999pt,Chan:1999nk,
Chaichian:2000eh,Chung:2001mb} 
is its position space
counterpart. It is clear then that
the radiative correction $(\Delta \kappa)_\mu$ is regularization dependent
and (at least to one loop) finite. The ambiguity has been
described in~\cite{Chung:1999gg} as resulting from the non-invariance
under chiral transformations of the measure in the path integral.
We note that the papers that claim
to obtain a unique (and ``correct'') result by the application of a
particular method are implicitly using a regularization
that fixes the ambiguity. It remains to see whether some symmetry
(namely, gauge invariance) determines the value of $(\Delta
\kappa)_\mu$. We would also like to understand why the correction is
finite at one loop.

Gauge invariance is implemented at the quantum level by the
Ward identities\footnote{Strictly, the Ward identities need only be
imposed for the renormalized amplitudes. However, we separate the
discussion of the regularized and renormalized theory in order to
compare with the literature.}. The relevant ones for our problem read
\be
p_\mu \Pi_{\mbox{\tiny odd}}^{\mu\nu}(p) = p_\nu \Pi_{\mbox{\tiny
odd}}^{\mu\nu}(p) = 0 \, \label{WIorig}.
\ee
These identities are automatically fulfilled for any $K(p)$ due to
the presence of the antisymmetric tensor. So it seems that gauge
invariance has nothing to say here. This is actually what one would
expect since the integrated ${\cal L}^{\mbox{\tiny CS}}$ is gauge
invariant and hence allowed. However, Coleman and Glashow argued that
the combination of gauge invariance and analiticity forbids the
correction $(\Delta \kappa)_\mu$~\cite{Coleman:1999ti}. The argument
(adapted to the present  
problem) goes as follows: Consider the 1PI correlation function
$\Gamma_{\mu\nu}(p,q)$ of two photons and one insertion of the
interaction ${\cal L}^b$ and allow for the two photons to carry
different momenta ($p$ and $q$, respectively). The insertion must
then carry momentum $p+q$. Differentiating the Ward identities
analogous to~(\ref{WIorig}) with respect to $p$ with fixed $q$ and
using analiticity at $p=0$ one learns that $\Gamma_{\mu\nu}(p,q)$
has at least one factor of $p$. The same reasoning applied to $q$
shows that 
it also has at least one factor of $q$. Thus
$\Gamma_{\mu\nu}(p,q)$ is $O(pq)$ and in the limit $q\rightarrow p$
(\ie, the limit in which $b_\mu$ is a constant) we find that it is
$O(p^2)$ and then $K(0)=0$. The assumption of analiticity requires
that there are no massless charged fermions (as is the case in
nature). Coleman and Glashow showed that the result also holds in the
presence of internal photons in multiloop calculations. Furthermore,
the same argument can be applied to the case in which there are more
than one CPT violating insertions (in this case, the diagram is
power-counting superficially convergent).

Even though this argument is perfectly correct, it rests on one
implicit requirement: gauge invariance is imposed for non-constant
$b_\mu$, \ie, when $b_\mu$ is promoted to a field $B_\mu(x)$. It is no
wonder that this prevents the induction of a Chern-Simons
term proportional to $b_\mu$, since $B\wedge A\wedge dA$ is not gauge
invariant. This strong requirement is not
necessary in the original formulation of the theory, in which $b_\mu$
is a constant. There,
the only relevant Ward identities are~(\ref{WI}), which do not
constrain the value of $(\Delta \kappa)_\mu$. Actually, the same
argument at the classical level would imply $\kappa_\mu=0$, but it is
clear that $\kappa_\mu \not =0$ is allowed by gauge invariance of the
action~\cite{Jackiw:1999yp}. Coleman and Glashow explicitly assumed gauge
invariance at the Lagrangian level, but their argument
is nevertheless useful for our problem. Indeed, it shows that
\begin{itemize}

\item $(\Delta \kappa)_\mu=0$ to all orders in $e$ (the electromagnetic
coupling) and $b_\mu$ in
regularization methods that preserve 
gauge invariance for non-constant $b_\mu$ and in which
$\Pi_{\mbox{\tiny odd}}^\mu(p) = \lim_{q\rightarrow -p}
\Gamma^{\mu\nu}(p,q)$. These include dimensional
regularization (as shown explicitly in \cite{Bonneau:2001ai}) and
Pauli-Villars\footnote{Note that there are methods preserving gauge
invariance and giving a non-vanishing $(\Delta \kappa)_\mu$. One example
is differential regularization, in which certain scales are fixed
by hand in such a way that the Ward identities are
satisfied~\cite{Freedman:1992tk}. Since one only needs to impose gauge
invariance in the original theory, any value can be
obtained~\cite{Chen:1999ws}. The result in constrained differential
renormalization~\cite{delAguila:1998kw,delAguila:1999nd} is not fixed
either~\cite{delAguila:1998gp}.}. 

\item The vacuum polarization in any other regularization procedure
can only differ, at one loop, by finite local terms. Therefore
$(\Delta \kappa)_\mu$ is one-loop finite in any
regularization~\cite{Bonneau:2001ai}. 

\item The only possible non-vanishing contribution at one loop is
linear in $b_\mu$. Indeed, higher order contributions are
power-counting finite. Since the Coleman-Glashow limit necesarily
coincides with the constant $b_\mu$ result for these convergent
diagrams, they must vanish.

\end{itemize}
The fact that the radiative correction to ${\cal L}^{\mbox{\tiny
CS}}$---a renormalizable term allowed by the symmetries---is
regularization dependent is not surprising. The only unusual
phenomenon here is that it is finite (at least at one loop). The
reason for this will become clearer below, when we discuss the relation
of this term with the ABJ anomaly.

To understand the significance of
an ambiguous finite correction\footnote{Other similar examples of
finite ambiguous radiative corrections are discussed
in~\cite{Jackiw:2000qq}. An example of a finite high-energy model of
Lorentz and CPT violation 
leading to a determined $(\Delta \kappa)_\mu$ was provided
in~\cite{Volovik:1999up}.}  
we need to go beyond simple regularization and discuss the 
renormalization of the theory. The renormalization of QED with the
CPT-violating terms ${\cal L}^{\mbox{\tiny CS}}$ and ${\cal L}^b$ has
been studied in detail by Bonneau in~\cite{Bonneau:2001ai}. However,
Bonneau considers a modification of the theory with sources for
the CPT-violating terms. As we have discussed, this is too restrictive
and leads to Ward identities that, combined with analiticity, forbid
the correction $(\Delta \kappa)_\mu$. We have argued that this is not
necessarily so in the theory ${\cal L}^{\mbox{\tiny QED}} + {\cal
L}^{\mbox{\tiny CS}} + {\cal L}^b $. To renormalize this theory we
must add the counterterms allowed by the symmetries. In
particular, we have
\be
{\cal L}^{\mbox{\tiny CS}}_R  =  \frac{1}{2}
(\kappa_\mu+ (\delta \kappa)_\mu)
\epsilon^{\mu\nu\rho\sigma} A_\nu 
 F_{\rho\sigma} \label{renCSterm} \, , 
\ee
where $\kappa_\mu$ is now a renormalized parameter. The
renormalized Ward identities have the form~(\ref{WIorig}),
with $\Pi^{\mu\nu}(p)$ the renormalized vacuum polarization. They do
not fix the local part of $\Pi_{\mbox{\tiny odd}}^{\mu\nu}(p)$ nor the
coefficient of the Chern-Simons like term in the 
effective action, which reads
\be
\kappa_\mu^{\mbox{\tiny eff}}=(\kappa + \Delta \kappa + \delta \kappa)_\mu \, ,
\ee
The counterterms are determined when a particular consistent
regularization method is used---this fixes $(\Delta \kappa)_\mu$---and
adequate normalization conditions are imposed. We
follow~\cite{Bonneau:2001ai} and choose the on-shell condition
\bea
\kappa_\mu & = & 
- \frac{1}{2}\epsilon_{\mu\nu\rho\sigma}\left[\frac{\d}{\d p_\sigma}
\la A^\nu(p) A^\rho(-p)\ra \right]_{p=0}  \nn
& = & \kappa_\mu^{\mbox{\tiny eff}}   \, . \label{normcond}
\eea
The renormalized 1PI correlation function in this
equation includes the tree level contribution.
For any regularization scheme, the normalization condition
(\ref{normcond}) implies 
\be
(\delta \kappa)_\mu  =  - (\Delta \kappa)_\mu  \, .
\ee
The counterterm is one-loop finite.
The renormalized parameter $\kappa_\mu$ is directly related to an
observable and can only be determined by the experiment.
We see that any regularization method is physically equivalent. 
In practice, a method like dimensional regularization is convenient
because it allows to set $(\delta \kappa)_\mu=0$ to all
orders~\cite{Bonneau:2001ai}. In the language of multiplicative renormalization the
parameter $\kappa_\mu$ is then renormalized like
$(\kappa_0)_\mu= Z_A^{-1} \kappa_\mu$, where 
$(\kappa_0)_\mu$ is a bare coupling and $Z_A^{1/2}$ is the wave function
renormalization of the photon. 

We are now in a position to answer the initial question about
possible restrictions on $b_\mu$. 
The observations indicate a vanishing or extremely small
$\kappa_\mu^{\mbox{\tiny eff}}$. According to (\ref{normcond}) we should
set $\kappa_\mu \approx 0$. Since $\kappa_\mu$ is an independent parameter
it can be set to zero for any value of $b_\mu$ (and any
regularization). Therefore, from the point of view of perturbative
renormalization theory we find no constraint on $b_\mu$. To constrain
$b_\mu$ we also need to measure $\la A^\nu(p)
A^\rho(-p)\ra_{\mbox{\tiny odd}}$ off
shell. There is a problem of naturalness though: if CPT and Lorentz
symmetries are broken, it is not explained why we should find an
(almost) vanishing value for $\kappa_\mu$. Even if the breaking of CPT
and Lorentz is suppressed, one needs a large amount of fine-tuning to
obtain values of $\kappa_\mu$ and $b_\mu$ that differ by many orders
of magnitude. Of course, these conclusions do not depend on the
normalization condition~(\ref{normcond}). To illustrate this
well-known fact, suppose that, at some order in perturbation theory,
$(\Delta \kappa)_\mu=\beta b_\mu=$ {\em finite} in a given
regularization 
and choose as normalization conditions minimal substraction, \ie,
substract only the divergences. Then, $(\delta \kappa)_\mu=0$ and
$\kappa_\mu^{\mbox{\tiny eff}}=\kappa_\mu+\beta b_\mu$. The
experiment tells us that $\kappa_\mu=-\beta b_\mu$. Although the
value of the renormalized parameter $\kappa_\mu$ is different, the
physical predictions are unchanged. Again, we see that there is a
fine-tuning.  

It is possible that the fine-tuning is explained by a
particular mechanism of symmetry breaking. Following Colladay and
Kosteleck\'y, we assume that CPT and Lorentz are spontaneously
broken. This occurs when
fields transforming non-trivially under these symmetries acquire
non-vanishing vacuum expectation values (vevs). 
Sponteanous symmetry breaking has the advantage of
preserving  observer Lorentz
invariance~\cite{Colladay:1998fq,Colladay:1997iz,Kostelecky:1999rh,
Colladay:2000jh,Colladay:1998pc,Kostelecky:2000mm}.  
We need to know more details about the high energy
theory. The simplest situation is that both
$b_\mu$ and $\kappa_\mu$ are generated as the vev of an axial vector field
$B_\mu$~\cite{Kostelecky:2000mm}. We can write 
\be
B_\mu(x)=B^{(1)}_\mu(x)+B^{(2)}_\mu(x)= \rho(x) \d_\mu \sigma(x) +
\d_\mu \phi(x)\, .  \label{decomposition}
\ee
We have argued above that gauge invariance forbids a term
$B\wedge A\wedge dA$ in the effective action. Therefore, as long as
$B_\mu(x)$ is treated as a single entity, no $\kappa_\mu$ can be
generated. But one could also renormalize $\phi$ and
$B^{(1)}$ independently. The term $d \phi \wedge A\wedge dA$ is gauge
invariant up to a total derivative. Hence it is allowed in the
classical action and radiative corrections can in principle contribute
to it. Upon spontaneous symmetry breaking $\la \d_\mu \phi \ra =
\mbox{constant}$, non-vanishing $b_\mu$ and $\kappa_\mu$ would be
generated. Similarly, one can also consider a model in which a
gauge-singlet real pseudoscalar $\phi$ (an axion) acquires a
non-constant vev of this kind\footnote{This mechanism was considered
in detail in~\cite{Andrianov:1999ay,Andrianov:1995qv}, where 
it was shown that a spacelike nonzero $\la \d_\mu \phi \ra$ can be
dynamically generated by radiative corrections.}. In the following we
study the radiative corrections for this model and then take $\la \phi
\ra \propto b\cdot x$. We can also regard this procedure as a
gauge-invariant modification of the Coleman-Glashow argument: we
promote $b_\mu$ to a 
non-constant field in such a way that an induced gauge-invariant
Chern-Simons like term is allowed, and then take the limit of constant
$b_\mu$. 

Both motivations lead us to study the Lorentz and CPT invariant
effective theory
\be
{\cal L}^\phi \supset \bar{\psi} (i\not \! D -m +
\frac{\tilde{b}}{\Lambda} \gamma_5 \not \! \d \phi + i c \gamma_5
\phi + \frac{d}{\Lambda}\phi^2) \psi -  \tilde{\kappa}
\frac{1}{2\Lambda} \phi F^{\mu\nu} \tilde{F}_{\mu\nu} \, , \label{action}
\ee
where $\tilde{F}_{\mu\nu} = 1/2 \,
\epsilon_{\mu\nu\rho\sigma}F^{\rho\sigma}$, $\Lambda$ is some large
scale and the last term is gauge 
invariant and equal to $\frac{\tilde{\kappa}}{2\Lambda}
\epsilon^{\mu\nu\rho\sigma} 
\d_\mu \phi A_\nu F_{\rho\sigma}$ up to a total derivative. The axion
$\phi$ is a gauge singlet and it can be external or
dynamical. We shall interpret this theory in two different ways: as a particular
model leading to CPT breaking and as an intermediate step to study the
original theory ${\cal L}^{\mbox{\tiny QED}}
+ {\cal L}^{\mbox{\tiny CS}} + {\cal L}^b$, which is recovered for $c=d=0$
and $\la \phi \ra=\frac{\Lambda}{\tilde{b}} b\cdot x$. We
shall work at order $1/\Lambda$ and, unless otherwise specified, we
assume $m\not \!=0$. 

We are interested in the radiative correction to
the coefficient of the last term in the effective action, which is given by 
\be
\Delta \tilde{\kappa} = -\frac{\Lambda}{2} C(0,0)\, \label{tildekappa}
,  
\ee
where
\be
\Gamma^{\mu\nu}(p,q)=\epsilon^{\mu\nu\rho\sigma}
p_\rho q_\sigma C(p,q)\, . \label{gamma} 
\ee
Here, $\Gamma^{\mu\nu}(p,q)$ is the radiative correction to the 1PI
correlation function of one 
axion and two photons, $\la A^\mu(p) A^\nu(q) \phi(-p-q) \ra$, and all
momenta are incoming. The form~(\ref{gamma}) is dictated by Lorentz
covariance. The Ward identities
\be
p_\mu \Gamma^{\mu\nu}(p,q) = q_\nu \Gamma^{\mu\nu} = 0 \, \label{WI}.
\ee
are automatically satisfied. Note that the application of
Coleman-Glashow argument tells us that $\Gamma^{\mu\nu}(p,q)\sim
O(pq)$, which in this case is 
obvious from the structure~ (\ref{gamma}). The corresponding
correction to the effective action can be nevertheless non-vanishing
because both momenta correspond to the derivatives in the interaction
of the generalized Chern-Simons term.

From power counting and Lorentz covariance, the contribution to
$\Gamma^{\mu\nu}$ linear in $c$ is superficially
convergent. A simple calculation gives
\be
(\Delta \tilde{\kappa})_c= \frac{ce^2 \Lambda}{8\pi^2 m}  \label{ccontrib} 
\ee
at one loop. The contribution proportional to $\tilde{b}$ is at most
(superficially) linearly divergent by power-counting, but again
the degree of divergence is lowered. A formal one-loop calculation
without any explicit regularization gives 
\bea
\lefteqn{\left[\frac{\d}{\d p_\rho} \frac{\d}{\d q_\sigma} \Gamma_b^{\mu\nu}(p,q)
\right]_{p=q=0}} && \nn  && =  \left[\frac{\d}{\d p_\rho} \frac{\d}{\d
q_\sigma} \left(2 \frac{\tilde{b}}{\Lambda} e^2 \int \frac{{\mathrm d}^4 k}{(2\pi)^4}
tr\{\gamma^\mu \frac{1}{\sla k -m} \gamma^\nu \frac{1}{\sla k -\sla q -m} 
(\sla p + \sla q) \gamma_5 
\frac{1}{\sla k + \sla q -m} \}\right) \right]_{p=q=0} \nn
&& = 16 i \frac{\tilde{b}}{\Lambda} e^2 \epsilon^{\mu\nu\rho\alpha}
\int \frac{{\mathrm d}^4 
k}{(2\pi)^4} \frac{4k^\sigma k_\alpha - (k^2-m^2)
\delta^\sigma_\alpha}{(k^2-m^2)^3} \nn
&& = - \frac{\tilde{b}e^2}{2 \pi^2 \Lambda} \epsilon ^{\mu\nu\rho\sigma}
\, . \label{naive}
\eea
The factor 2 in the second line corresponds to the contributions of
the two possible momentum flows, which are identical after a shift
in $k_\mu$. In the last line we have used symmetric integration: 
\be 
\int {\mathrm d}^4 k k^\sigma k_\alpha f(k^2) =  \frac{1}{4}
\delta^\sigma_\alpha \int {\mathrm
d}^4 k k^2 f(k^2) \label{symmid}.
\ee
Hence we find $(\Delta \tilde{\kappa})_{\tilde{b}}= \tilde{b}e^2 /
4\pi^2$. Taking $\phi(x)=\frac{\Lambda}{\tilde{b}} b \cdot x$ and
$c=0$ we would arrive at $(\Delta \kappa)_\mu = e^2 / 4\pi^2 \,
b_\mu$. However, some of the identites we have used in~(\ref{naive}) are not
well-defined for divergent 
integrals. A rigorous calculation requires the use of some
regularization procedure and the result can differ from the one above.
In particular, (\ref{symmid}) does not hold in general. Moreover,
depending on what regularization one uses,
there may be additional contributions coming from the shift in $k$ or
from the Dirac algebra. We note in passing that if we add to the
result above the contribution of the momentum shift,
following~\cite{Jackiw:1999yp}, we obtain 
$(\Delta \kappa)_\mu = 3 e^2 / 16\pi^2 b_\mu$. This is the result
found in non-perturbative calculations in $b_\mu$ that respect the
identity~(\ref{symmid})~\cite{Jackiw:1999yp,
Perez-Victoria:1999uh,Chung:1999pt}.   

In order to discuss the regularization dependence and the possible
restrictions of gauge invariance it is useful to write the
contribution linear in $\tilde{b}$ in terms of the chiral triangle:
\be
\Gamma_{\tilde{b}}^{\mu\nu}(p,q) =
\frac{\tilde{b}e^2}{\Lambda}(p_\lambda+q_\lambda)
V^{\mu\nu\lambda}(p,q) \, . \label{triangle}
\ee
$V^{\mu\nu\lambda}(p,q)=\la J^\mu(p) J^\nu(q) J_5^\lambda(-p-q)\ra$ is
the correlator of one chiral current $J_5^\lambda=\bar{\psi} \gamma_5
\gamma^\lambda \psi$ and two vector currents $J^\mu=\bar{\psi}
\gamma^\mu \psi$. The theory at hand has an invariance under chiral
transformations of the fermion fields (not acting on $\phi$) which is
broken explicitly by the fermion mass term and also by the ABJ
anomaly. This leads to the anomalous Ward
identity 
\be
(p_\lambda+q_\lambda) V^{\mu\nu\lambda}(p,q) = 
2m i V^{\mu\nu}(p,q) + \epsilon^{\mu\nu\rho\sigma} p_\rho q_\sigma \,
{\cal A}  \, \label{chiralWI}
\ee
where $V^{\mu\nu}(p,q)=\la J^\mu(p) J^\nu(q) J_5(-p-q)\ra$, with
$J_5=\bar{\psi}\gamma_5\psi$, and ${\cal A}$ is the anomaly, which is
independent of $p,q$ (it is a delta function in position space).
$V^{\mu\nu}(p,q)=\epsilon^{\mu\nu\rho\sigma}p_\rho q_\sigma V(p,q)$ is
finite, regulator independent and essentially 
equal to the contribution to $\Gamma^{\mu\nu}$ linear in $c$.
${\cal A}$ is also finite but it is regularization dependent. However,
if the Ward identities for the vector currents,
\be
p_\mu V^{\mu\nu\lambda}(p,q)=q_\nu V^{\mu\nu\lambda}(p,q)=0 \,
\label{vectorWI}, 
\ee
are enforced, the value of the chiral
anomaly is fixed to be ${\cal A}=\frac{1}{2\pi^2}$.
From~(\ref{gamma}), (\ref{triangle}) and (\ref{chiralWI}) we find
\be
C_{\tilde{b}}(p,q)= \frac{\tilde{b}e^2}{\Lambda} (2m V(p,q)+{\cal A})
\ee
and from~(\ref{tildekappa}) and (\ref{ccontrib}),
\be
(\Delta \tilde{\kappa})_b = \tilde{b}e^2 (\frac{1}{4\pi^2} - \frac{1}{2}
{\cal A}) \, , 
\ee
Taking $\la \phi \ra =\frac{\Lambda}{\tilde{b}} b \cdot x$ and $c=0$
we obtain 
\be
(\Delta \kappa)_\mu = e^2 b_\mu (\frac{1}{4\pi^2} - \frac{1}{2} {\cal
A}) \, , \label{generalkappa}
\ee 
If the equations~(\ref{vectorWI}) are satisfied we find
$\Delta \tilde{\kappa} =0$. This can be understood in the following
way~\cite{Frishman:1981dq}. Identity~(\ref{chiralWI}) and
analiticity in $p$ and $q$ imply that  
\be
V^{\mu\nu\lambda}(p,q)=\frac{1}{2}(2mV(0,0)+{\cal
A})\epsilon^{\mu\nu\lambda\rho} (p_\rho-q_\rho) (1+O(p)+O(q))
\ee
The identities~(\ref{vectorWI}) then require $2mV(0,0)+{\cal
A}=0$. 
In many applications, the vector Ward
identities~(\ref{vectorWI}) correspond to gauge invariance and must
be preserved. One might then think that gauge invariance determines 
$\Delta \tilde{\kappa} = 0$ and, for the original problem, $(\Delta
\kappa)_\mu = 0$. This is not right because in this case gauge
invariance does not imply~(\ref{vectorWI}) but only
\be
p_\mu (p_\lambda+q_\lambda) V^{\mu\nu\lambda}(p,q)=q_\nu
(p_\lambda+q_\lambda) V^{\mu\nu\lambda}(p,q)=0 \,
\label{modvectorWI}\, .
\ee
Of course, the identities (\ref{modvectorWI}) are nothing but the Ward
identities~(\ref{WI}), and are satisfied for any value of ${\cal
A}$. Since in our model the photon does not couple to any chiral
current, there is no fundamental reason for imposing
(\ref{vectorWI}) in this case. Therefore $(\Delta \kappa)_\mu$ is
also regularization dependent in our method.
Equation (\ref{generalkappa}) shows explicitly the relation
between the induced Chern-Simons term and the chiral anomaly in
perturbation theory, and explains several facts observed in the
literature: 
\begin{itemize}

\item The radiative correction is finite to one-loop because it
is the sum of a finite amplitude and the chiral anomaly. The anomaly
accounts for the ambiguity.

\item The induced term vanishes at one loop in regularization
methods that respect the Ward identities~(\ref{WI}), like dimensional
regularization or Pauli-Villars. We stress again that this does not
mean that gauge invariance requires a vanishing correction. In order
to extend this result to higher loops one would need to show that
analiticity is not lost, as in the proof of Coleman-Glashow theorem.  

\item $(\Delta \kappa)_\mu$, though independent of $m$ for $m\not =0$,
is discontinuos at $m=0$ (in perturbation theory about
$b_\mu=0$). Indeed, in the massless case $V(p,q)=0$  
for dimensional reasons (and Lorentz covariance), so the first term of
(\ref{generalkappa}) is absent and the radiative correction is purely
local. The difference 
between the massive and massless case for any fixed regularization is
\be
(\Delta \kappa)^m_\mu-(\Delta \kappa)^0_\mu= \frac{e^2}{4\pi^2}  b_\mu
\, , 
\ee
which agrees with the calculation
in~\cite{Perez-Victoria:1999uh}. (There it was shown 
that the exact result for a spacelike $b_\mu$ is continuous in $m$ at
$m=0$; the equation above remains valid if $|b^2|<m^2$.) 
In perturbation theory the origin of the 
discontinuity is a potential infrared divergence (that also cancels 
out) when $m=0$ and $p=q=0$. Note that for $m=0$ 
dimensional regularization gives a non-vanishing result. 
This does not contradict our argument above nor Coleman-Glashow
theorem: the assumption of analiticity is invalidated by the presence
of massless fermions. 

\item Higher orders in $b_\mu$ would arise from correlation functions
of two photons and more than one axion. The generalization of
(\ref{triangle}) for these correlators can be studied in a
similar way. The corresponding diagrams are superficially convergent
and analiticity in the external momenta can be used to
show that higher orders in $b_\mu$ do not contribute. In the
massless case this is straightforward by dimensional analysis.

\end{itemize}
The same connection with the anomaly was established in
the path integral formalism in~\cite{Chung:1999gg}. That method
can also be employed in the axion model: One can
eliminate the term proportional to $\tilde{b}$ by a local chiral
transformation of the fermion fields with the parameter of the
transformation proportional to the axion. This redefines $c$, $d$ and
$\tilde{\kappa}$. The redefinition of $\tilde{\kappa}$, arising from
the Fujikawa Jacobian, obviously corresponds
to the anomaly in~(\ref{chiralWI}), whereas the radiative correction
induced by the new part of $c$ accounts for the first term on the
\rhs\ of~(\ref{chiralWI}).

Let us consider now the renormalization of the
theory~(\ref{action}). It is important to note that we are in a
decoupling scenario~\cite{Appelquist:1975tg,Wudka:1994ny}, so in
perturbation theory there 
are two expansion parameters: $e$ and $\Lambda$. We want
renormalization to 
preserve the expansion in $\Lambda$. The radiative correction has
the form
\be
\frac{\Delta \tilde{\kappa}}{\Lambda}= c O(e^2) +
\tilde{b} O(e^2) \frac{1}{\Lambda} + O(e^2 \frac{1}{\Lambda^2}) \, .
\ee
While the correction $\tilde{b} O(e^2)$ can be absorbed by a
counterterm $\delta \tilde{\kappa}$, no counterterm with the adequate
power in $\Lambda$ can absorb the correction linear in $c$. Hence,
there is a 
definite finite contribution given, at order $c e^2$,
by~(\ref{ccontrib}). This is analogous to the case of $g-2$ in QED,
which is not an independent parameter but a prediction of the theory
(non-renormalizable operators describing new physics can only
give small corrections to the QED value).
With appropiate normalization
conditions\footnote{The normalization condition for $\tilde{\kappa}$
should be imposed for the contribution of order $1/\Lambda$ to the
corresponding correlation function.},  
\be
\frac{\tilde{\kappa}^{\mbox{\tiny eff}}}{\Lambda}=
\frac{ce^2}{8\pi^2 m}  +
\frac{\tilde{\kappa}}{\Lambda} + \mbox{higher orders} \,
. \label{general} 
\ee
If $\d_\mu \phi$ gets a vev we recover the original
theory ${\cal L}^{\mbox{\tiny QED}} + {\cal L}^{\mbox{\tiny
CS}} + {\cal L}^b$ plus extra terms proportional to $c$ and $d$ and
those describing the excitations about this
vacuum. (\ref{general}) translates into an analogous equation for
$\kappa_\mu^{\mbox{\tiny eff}}$. In particular, for non-zero $c$, a
Chern-Simons term not suppressed by $\Lambda$ is induced.   
The strong limits on $\kappa_\mu^{\mbox{\tiny eff}}$ then imply
strong limits on $c$. In fact, not only Lorentz and CPT but also
translational invariance is broken in the fermion sector unless
$c=d=0$. We would have $c=d=0$ automatically if the axion were a
Goldstone boson of some 
spontaneously broken symmetry, as in the axion solution of the strong
CP problem~\cite{Peccei:1977hh,Weinberg:1978ma,Wilczek:1978pj}. 
On the other hand the pure photon sector puts no bounds on
$\tilde{b}$, but $b_\mu \gg \kappa_\mu$ requires a fine-tuning of the
high-energy parameters, $\tilde{b} \gg \tilde{\kappa}$.

Finally, let us consider again the model in which $b_\mu$ is the vev
of a vector boson and use the decomposition~(\ref{decomposition}).
We have just studied the behaviour of $B^{(2)}$. On the other hand,
the term $B^{(1)}\wedge A\wedge dA $ in the effective action is
forbidden by gauge invariance. 
If $\la \phi \ra =0$ but $B^{(1)}$ gets a non-vanishing
(small) vev, we obtain a theory with $\kappa_\mu=0$ to all orders and
$b_\mu \not = 0$. This would account for a possible sizable CPT and
Lorentz violation in the fermion sector compatible with the limits from the
absence of birefringence of the light. Observe that we have not used
anomaly cancelation in the fundamental theory as in~\cite{Colladay:1998fq} 
(although $B_\mu$ could be in particular an extra gauge field). We
will not try here to propose a realistic field-theoretical model of
CPT and Lorentz violation. As a matter of fact, stability and
causality at all scales seem to require non-local
physics~\cite{Kostelecky:2000mm}. We should also mention here an
interesting alternative to spontaneous Lorentz and CPT breaking.
In~\cite{Klinkhamer:2000zh}, Klinkhamer has shown that quantum effects
can generate a non-zero $(\Delta \kappa)_\mu$ in a classically Lorentz
and CPT invariant chiral gauge theory when at least one
dimension is compact. This CPT anomaly arises when a non-perturbative
gauge anomaly is cancelled, similarly to the case of the
parity anomaly in certain three-dimensional gauge theories. In this
scenario the size of the Chern-Simons term is a definite prediction of the
theory, fixed by gauge invariance and the classical Lorentz and CPT
symmetries. Furthermore, this term is naturaly very small, as
it is inversely proportional to the size of the universe.

Let us summarize our conclusions. $\cal{L}^{\mbox{\tiny CS}}$ is a
local renormalizable term allowed by the symmetries of the theory
once Lorentz and CPT have been broken. Therefore its coefficient in
the effective action is an independent parameter and can be
determined only by comparison with the experiment. At this level the
observed smallness of this coefficient is compatible with a larger CPT
violation in the fermion sector. This 
entails, however, a fine-tuning that can be explained by
particular field theory models.
Using the theory~(\ref{action}) we have also obtained a relation
between the induced CPT violating Chern-Simons term and the ABJ
anomaly which accounts for the (one-loop) convergence of the radiative 
corrections. 

\vspace*{1cm}

\centerline{\bf Acknowledgments} 
It is a pleasure to thank R. Jackiw for enlightening discussions. I
thank MECD for financial support.

\end{document}